\documentclass{article}

\usepackage{iclr2026_conference,times}

\iclrfinalcopy

\usepackage{hyperref}
\usepackage{url}
\usepackage{amsmath,amsfonts,amssymb}
\usepackage{graphicx}
\usepackage{booktabs}
\usepackage{xcolor}
\usepackage{float}

\title{Biologically-Grounded Multi-Encoder Architectures as Developability Oracles for Antibody Design}

\author{Simon J. Crouzet\thanks{Corresponding author: \texttt{simon.crouzet@gmail.com}}}

\begin{document}

\renewcommand{\footrulewidth}{0.4pt}
\fancyfoot[C]{\small ICLR 2026 Workshop on Generative and Experimental Perspectives for Biomolecular Design\\[0.2ex]
Camera-ready -- 15 April 2026 \quad \href{https://openreview.net/forum?id=UPUoa6mcdZ}{\texttt{openreview.net/forum?id=UPUoa6mcdZ}}}

\maketitle

\begin{abstract}
Generative models can now propose thousands of \emph{de novo} antibody sequences, yet translating these designs into viable therapeutics remains constrained by the cost of biophysical characterization. Here we present CrossAbSense, a framework of property-specific neural oracles that combine frozen protein language model encoders with configurable attention decoders, identified through a systematic hyperparameter campaign totaling over 200 runs per property. On the GDPa1 benchmark of 242 therapeutic IgGs, our oracles achieve notable improvements of 12--20\% over established baselines on three of five developability assays and competitive performance on the remaining two. The central finding is that optimal decoder architectures \emph{invert} our initial biological hypotheses: self-attention alone suffices for aggregation-related properties (hydrophobic interaction chromatography, polyreactivity), where the relevant sequence signatures --- such as CDR-H3 hydrophobic patches --- are already fully resolved within single-chain embeddings by the high-capacity 6B encoder. Bidirectional cross-attention, by contrast, is required for expression yield and thermal stability --- properties that inherently depend on the compatibility between heavy and light chains. Learned chain fusion weights independently confirm heavy-chain dominance in aggregation ($w_H = 0.62$) versus balanced contributions for stability ($w_H = 0.51$). We demonstrate practical utility by deploying CrossAbSense on 100 IgLM-generated antibody designs, illustrating a path toward substantial reduction in experimental screening costs.
\end{abstract}

\section{Introduction}

Therapeutic monoclonal antibodies constitute a pharmaceutical market exceeding \$250 billion, yet approximately 30\% of clinical-stage candidates exhibit biophysical liabilities --- aggregation, poor expression, or thermal instability --- that compromise manufacturability and ultimately limit clinical success~\citep{Jain2017}. Maintaining acceptable quality attributes throughout development remains a central challenge in biopharmaceutical manufacturing~\citep{Schiestl2011}, and the emergence of generative models capable of proposing thousands of \emph{de novo} antibody sequences~\citep{Shuai2023IgLM,Dreyer2025} has only sharpened the need for rapid, reliable computational pre-filtering. Without robust in-silico developability oracles, the vast majority of generated candidates cannot be experimentally assessed, creating a critical bottleneck between computational design and therapeutic reality.

Computational approaches to developability prediction have advanced considerably, from hand-crafted sequence profiling tools~\citep{Raybould2022TAP} to machine learning methods that jointly optimize binding affinity and biophysical properties~\citep{Makowski2024}. The mechanistic heterogeneity of developability --- spanning surface hydrophobicity~\citep{Lee2013,Raybould2019TAP}, electrostatic self-association~\citep{Chaudhri2013,Dobson2016}, VH--VL pairing efficiency~\citep{Jayaram2012}, and cooperative domain stability~\citep{Guo2015} --- presents a unique opportunity. By allowing a neural architecture search to select which computational strategy best predicts each property, we can probe which biophysical signals are encoded directly in sequence (and likely shaped by evolution), and which depend on structural interactions between chains that no single-chain representation can capture.

Here we exploit this opportunity by designing CrossAbSense --- our framework of property-specific neural oracles with three attention strategies that differ in how they model the relationship between heavy and light chains: (i)~\emph{self-attention only}, which processes each chain independently, reading the signal entirely from per-chain sequence features; (ii)~\emph{self+cross attention}, which first builds intra-chain context then queries across chains, analogous to a fold-then-assemble pathway; and (iii)~\emph{bidirectional cross-attention}, which allows each chain to continuously query the other, explicitly modeling the paired interface. A systematic hyperparameter campaign, spanning encoder selection, decoder architecture, sequence representation, and training configuration, discovers which strategy best predicts each of five GDPa1 benchmark assays~\citep{Arsiwala2025}. The results reveal that aggregation-related properties are predicted best by per-chain reasoning alone, while expression and thermal stability require explicit inter-chain modeling: an inversion of our initial biological expectations with implications for how these properties are encoded in antibody sequence.

\section{Methods}

We evaluate on the GDPa1 benchmark~\citep{Arsiwala2025}, which provides measured values for 242 therapeutic IgGs across five developability assays: hydrophobic interaction chromatography (HIC), affinity-capture self-interaction nanoparticle spectroscopy (AC-SINS), polyreactivity in CHO lysate (PR\_CHO), expression titer (Titer), and CH2 domain thermal stability (Tm2). All experiments use 5-fold cross-validation with hierarchical clustering and IgG isotype stratification, ensuring that sequence-similar antibodies are separated across folds.

Each antibody chain is encoded using ESM-Cambrian~\citep{ESM_Science2024}, a general-purpose protein language model from the ESM family~\citep{Rives2021}, tested in 300M, 600M, and 6B parameter variants. ProtT5~\citep{Elnaggar2022ProtTrans}, a T5-based encoder pre-trained on UniRef50, was also evaluated as an alternative general-purpose encoder. We encode full heavy and light chain sequences (including variable and constant regions) --- a natural match for these encoders, which are pre-trained on full-length protein sequences and thus represent full chains within their learned distribution. Encoders remain frozen throughout training, preserving pre-trained evolutionary and structural knowledge while reducing trainable parameters by two orders of magnitude. Antibody-specific language models~\citep{Ruffolo2021AntiBERTy,Olsen2024} and structure-enhanced variants~\citep{Barton2024} were also evaluated during the encoder selection phase.

The decoder processes per-chain embeddings through $L$ pre-normalized attention layers with hidden dimension $d_h$, residual connections, and feed-forward blocks (expansion factor~4). We define three attention strategies, each encoding a different hypothesis about how biophysical information is distributed across the two chains:
\begin{itemize}
    \item \emph{Self-attention only}: each chain attends exclusively to its own residues across all $L$ layers; heavy and light chains are processed independently, testing whether the property signal can be read entirely from per-chain sequence features.
    \item \emph{Self\,+\,cross attention}: each layer first applies intra-chain self-attention, then inter-chain cross-attention where the heavy chain queries light-chain residues and vice versa. This mimics a fold-then-assemble pathway in which each chain consolidates its own representation before querying its partner.
    \item \emph{Bidirectional cross-attention}: each layer applies only cross-attention (the heavy chain queries light-chain residues and the light chain queries heavy-chain residues) without any intra-chain self-attention. This explicitly models paired VH--VL interface compatibility and cooperative inter-chain signals.
\end{itemize}

After the attention layers, chain representations are pooled and fused via a learnable weight:
\begin{equation}
    \mathbf{h} = w_H \, \mathbf{h}_H + (1 - w_H) \, \mathbf{h}_L, \quad w_H = \sigma(\theta_w)
    \label{eq:fusion}
\end{equation}
where $\theta_w$ is a learned scalar parameter. This provides an interpretable, data-driven quantification of each chain's contribution to the predicted property. A multi-layer prediction head produces the final scalar output $\hat{y}$.

We swept encoder type (ESM-Cambrian 300M/600M/6B, AntiBERTy, ProtT5), sequence representation (variable-only, AHO-aligned, full-chain), attention strategy, architecture dimensions, antibody-specific structural features, training schedule, Stochastic Weight Averaging~\citep{Izmailov2018SWA}, and loss function using Bayesian optimization~\citep{Snoek2012} with Hyperband early termination~\citep{Li2018Hyperband}. In total, over 200 configurations were evaluated per property, each under full 5-fold cross-validation, optimizing mean validation Spearman~$\rho$. We note that this campaign is, by design, an architectural comparison: the primary axis of variation is discrete --- encoder type, attention strategy, sequence representation --- not numerical fine-tuning. Furthermore, specialized developability benchmarks are inherently limited by the cost of experimental characterization; clustered cross-validation provides a strong generalization proxy in this regime.

\section{Results}

Table~\ref{tab:main_results} summarizes our results against seven representative baselines from the GDPa1 evaluation~\citep{Arsiwala2025}. We observe notable improvements on three properties: expression titer ($\rho = 0.428$, +20\% over previous best), thermal stability ($\rho = 0.387$, +18\%), and polyreactivity ($\rho = 0.475$, +12\%). Steiger's Z-test for dependent correlations~\citep{Steiger1980} yields $p < 0.02$ for all three under an assumed inter-model correlation of $r_{12} = 0.90$ --- a reasonable assumption given that models trained on the same data sources with overlapping feature representations tend to share systematic biases, succeeding and failing on the same cases. We nonetheless note that this assumption is difficult to verify empirically, and these p-values should be interpreted with caution given the small sample size ($N = 242$). On hydrophobic interaction chromatography and self-association, we achieve competitive performance within 2--7\% of the current best single-property specialists.

\begin{table}[t]
\centering
\caption{GDPa1 benchmark performance (Spearman $\rho$, 5-fold cluster-stratified CV). All baselines from~\citet{Arsiwala2025}. Best results per property are highlighted in \textbf{bold}.}
\label{tab:main_results}
\vspace{2pt}
\begin{tabular}{lccccc}
\toprule
\textbf{Method} & \textbf{HIC} & \textbf{AC-SINS} & \textbf{PR\_CHO} & \textbf{Titer} & \textbf{Tm2} \\
\midrule
TAP linear          & 0.222 & 0.294 & 0.136 & 0.113 & $-$0.115 \\
p-IgGen             & 0.346 & 0.388 & 0.424 & 0.238 & $-$0.119 \\
ESM2+Ridge          & 0.416 & 0.420 & 0.420 & 0.180 & $-$0.098 \\
ESM2+TAP+Ridge      & 0.420 & 0.480 & 0.413 & 0.221 & 0.265   \\
AbLang2             & 0.461 & \textbf{0.509} & 0.362 & 0.356 & 0.101   \\
MoE baseline        & \textbf{0.656} & 0.424 & 0.353 & 0.184 & 0.107   \\
DeepSP+Ridge        & 0.531 & 0.348 & 0.257 & 0.114 & 0.073   \\
\midrule
\textbf{CrossAbSense} & 0.644 & 0.475 & \textbf{0.475} & \textbf{0.428} & \textbf{0.387} \\
\bottomrule
\end{tabular}
\end{table}

The most informative outcome of our campaign is \emph{which} attention strategy the optimization selects for each property, and how these selections challenge our starting assumptions. We had hypothesized that aggregation-related properties --- hydrophobic interaction chromatography and polyreactivity --- would benefit from cross-attention, since aggregation was traditionally attributed to exposed patches at the VH--VL interface. Instead, self-attention alone proved optimal for both. With the high-capacity ESM-Cambrian 6B encoder --- whose approximately 120 internal attention layers already construct a rich per-residue structural context, the sequence signatures that drive aggregation, such as hydrophobic CDR-H3 motifs~\citep{Raybould2019TAP,Lee2013}, are already fully resolved within single-chain embeddings. The decoder does not need to query the partner chain: if a heavy chain carries an aggregation-prone motif, the risk is present regardless of which light chain it is paired with. Polyreactivity follows the same pattern, consistent with non-specific binding being driven by localized charge and hydrophobicity features on individual chain surfaces~\citep{Dobson2016}.

Conversely, expression titer and thermal stability both require bidirectional cross-attention. For expression titer, this aligns with a well-known biological principle: expression depends not merely on individual chain quality but on the efficiency of VH--VL heterodimerization and quaternary assembly~\citep{Jayaram2012}, where interface mutations can unpredictably alter binding kinetics~\citep{Khalifa2000}. Two individually well-folded chains may pair poorly, yielding low expression; the decoder must perform an explicit ``compatibility check'' between heavy and light chains to make accurate predictions.

The thermal stability result is perhaps the most thought-provoking. CH2 domain melting temperature is conventionally treated as a domain-intrinsic property, largely determined by isotype~\citep{Lee2013}. The model's preference for cross-attention raises the possibility that inter-chain coupling --- through disulfide bonds, VH--VL interface packing, and hinge-region mechanics --- modulates the cooperative thermal breathing of the whole molecule~\citep{Guo2015}. DSC studies have shown that variable domains contribute measurably to overall IgG1 thermal stability~\citep{Ionescu2008}, supporting the idea that stability is not purely a constant-region property. While this interpretation remains a hypothesis requiring experimental validation, it illustrates how architecture selection can generate testable mechanistic predictions.

We interpret this asymmetry through the lens of encoder capacity. ESM-Cambrian 6B appears to ``saturate'' local feature detection: with sufficient encoder capacity, all aggregation-relevant information is captured \emph{in cis}. But no single-chain encoder, however large, can represent inter-chain \emph{compatibility} --- that information is irreducibly bivariate. Because the decoder remains small relative to the encoder, it lacks the capacity to memorize shortcuts through unnecessary cross-attention paths; only genuinely informative inter-chain connections survive training. Its topology thus provides a window into the relational complexity of each biophysical property.

The learnable chain fusion weights (Eq.~\ref{eq:fusion}) offer an independent perspective on chain importance. For aggregation-related properties (hydrophobic interaction chromatography, self-association), training converges to $w_H = 0.62$, consistent with the known role of heavy-chain CDR-H3 hydrophobic patches as primary aggregation drivers~\citep{Raybould2019TAP,Chaudhri2013}. For thermal stability, $w_H = 0.51$, reflecting balanced chain contributions expected for a global molecular property. The convergence of two independent signals --- attention strategy selection and fusion weight learning --- toward the same conclusion strengthens the interpretation that aggregation is driven primarily by per-chain features, while stability and expression depend on both chains.

Full-chain encoding (VH+CH / VL+CL) outperforms variable-region-only representation on four of five properties, with 15--20\% improvement for thermal stability and expression titer. Constant regions encode IgG subclass identity (IgG1 vs.\ IgG4), providing biophysical baselines critical for stability and expression. The sole exception is self-association (AC-SINS), where the signal is localized to the paratope and benefits from the more focused Fv representation.

To demonstrate practical utility beyond benchmark performance, we generated 100 novel antibody designs using IgLM~\citep{Shuai2023IgLM} with the trastuzumab (Herceptin) framework as prompt --- chosen as one of the most extensively characterized therapeutic antibodies --- and scored them with our property-specific oracles (Table~\ref{tab:iglm_oracle}). The designs produce paired VH+VL sequences that explore CDR diversity while maintaining the human scaffold. Oracle predictions reveal that all 100 designs improve over trastuzumab on hydrophobic interaction chromatography, consistent with IgLM preserving the favorable hydrophobicity profile of its template while diversifying CDRs. However, none surpass trastuzumab on expression titer, self-association, or thermal stability; for polyreactivity, trastuzumab itself scores zero, so designs can at best match --- not beat --- the reference, and most do. Predicted values cluster in a narrow band (e.g., Titer standard deviation of 7.3\,mg/L across designs vs.\ 122.8\,mg/L across the training set; Figure~\ref{fig:iglm_delta} in Appendix). This lack of property diversity in unguided generation directly motivates using these oracles as reward functions for property-guided design. With inference throughput on the order of $10^4$ antibodies per day on a single GPU, these oracles can screen entire generative libraries at negligible computational cost.

\begin{table}[t]
\centering
\caption{Oracle predictions on 100 IgLM-generated trastuzumab-based designs. Train statistics from the GDPa1 benchmark (242 IgGs). ``\% beat Tras.''\ indicates fraction of designs predicted to improve over the trastuzumab reference (lower is better for HIC, PR\_CHO, AC-SINS; higher for Titer, Tm2). Note that trastuzumab scores zero on PR\_CHO, so designs can at best match the reference.}
\label{tab:iglm_oracle}
\vspace{2pt}
\resizebox{\textwidth}{!}{%
\begin{tabular}{lcccccc}
\toprule
\textbf{Property} & \textbf{Train mean} & \textbf{Train std} & \textbf{IgLM mean} & \textbf{IgLM std} & \textbf{Tras.\ ref} & \textbf{\% beat Tras.} \\
\midrule
HIC (min)      & 2.82  & 0.34   & 2.43  & 0.08  & 2.67   & 100/100 \\
Titer (mg/L)   & 240.6 & 122.8  & 186.3 & 7.3   & 352.4  & 0/100   \\
PR\_CHO (0--1) & 0.17  & 0.16   & 0.03  & 0.07  & 0.00   & 0/100   \\
AC-SINS (nm)   & 6.42  & 8.77   & 15.02 & 3.09  & 1.00   & 0/100   \\
Tm2 (\textdegree C)     & 82.16 & 3.01   & 78.83 & 0.89  & 82.75  & 0/100   \\
\bottomrule
\end{tabular}%
}
\end{table}

\section{Discussion}

Our results suggest that property-specific neural architectures can serve as more than predictive tools --- they offer a lens into the mechanistic structure of antibody developability. The finding that aggregation-related properties are best predicted by per-chain self-attention, while expression and stability require cross-attention, was not designed into the system but emerged from hyperparameter optimization. This carries direct implications for antibody engineering: aggregation liabilities can be addressed by optimizing individual chains in isolation, while expression and stability demand explicit VH--VL co-optimization.

More broadly, our findings suggest that the choice of decoder architecture --- when selected under parameter constraints by a systematic search --- can reveal whether a target property is driven by per-chain sequence features or by inter-chain structural relationships. For expression titer and thermal stability, relational reasoning through cross-attention is not merely beneficial but \emph{necessary} --- the relevant information is irreducibly bivariate. This principle may extend to other multi-chain or multi-domain protein systems where functional properties arise from inter-subunit coupling.

By operating entirely at the sequence level, our approach probes what protein language models have already learned about developability from evolutionary data alone. Incorporating three-dimensional structure or post-translational context may further improve predictions and remains a natural extension. Whether zero-shot PLM likelihoods or context-aware structure-based scores (e.g., inverse folding models) can serve as training-free developability proxies --- avoiding the need for labeled data entirely --- remains an open and promising direction. The GDPa1 benchmark, while rigorous, comprises only 242 IgGs; generalization to other antibody formats --- Fabs, scFvs, or nanobodies, where developability profiling is rapidly advancing~\citep{Gordon2026TNP} --- remains to be tested. The oracle experiment relies on in-silico predictions without wet-lab confirmation. These oracles nonetheless enable immediate integration into generative antibody design workflows, whether as differentiable reward models for reinforcement learning, classifier guidance for diffusion-based generators, or acquisition functions for adaptive experimental design. Our IgLM experiment illustrates this potential: the narrow clustering of unguided designs across developability space shows that generative models alone do not optimize for biophysical quality, whereas coupling them with property-specific oracles could steer sampling toward regions that balance sequence novelty with manufacturability. This positions developability prediction as essential infrastructure for the next generation of antibody therapeutics.

\subsubsection*{Acknowledgments}
We thank Ginkgo Bioworks for organizing the GDPa1 Developability Prediction competition and releasing the benchmark dataset that made this work possible~\citep{Arsiwala2025}. All code, model checkpoints, and evaluation scripts are available at \url{https://github.com/SimonCrouzet/CrossAbSense}. Claude (Anthropic) provided writing and styling assistance; all references were compiled manually.

\setlength{\bibsep}{4pt}
\bibliographystyle{iclr2026_conference}
\bibliography{PaperDraft_GEM@ICLR2026}

\appendix
\section{IgLM Oracle Validation}
\label{app:iglm}

\begin{figure}[H]
\centering
\includegraphics[width=\textwidth]{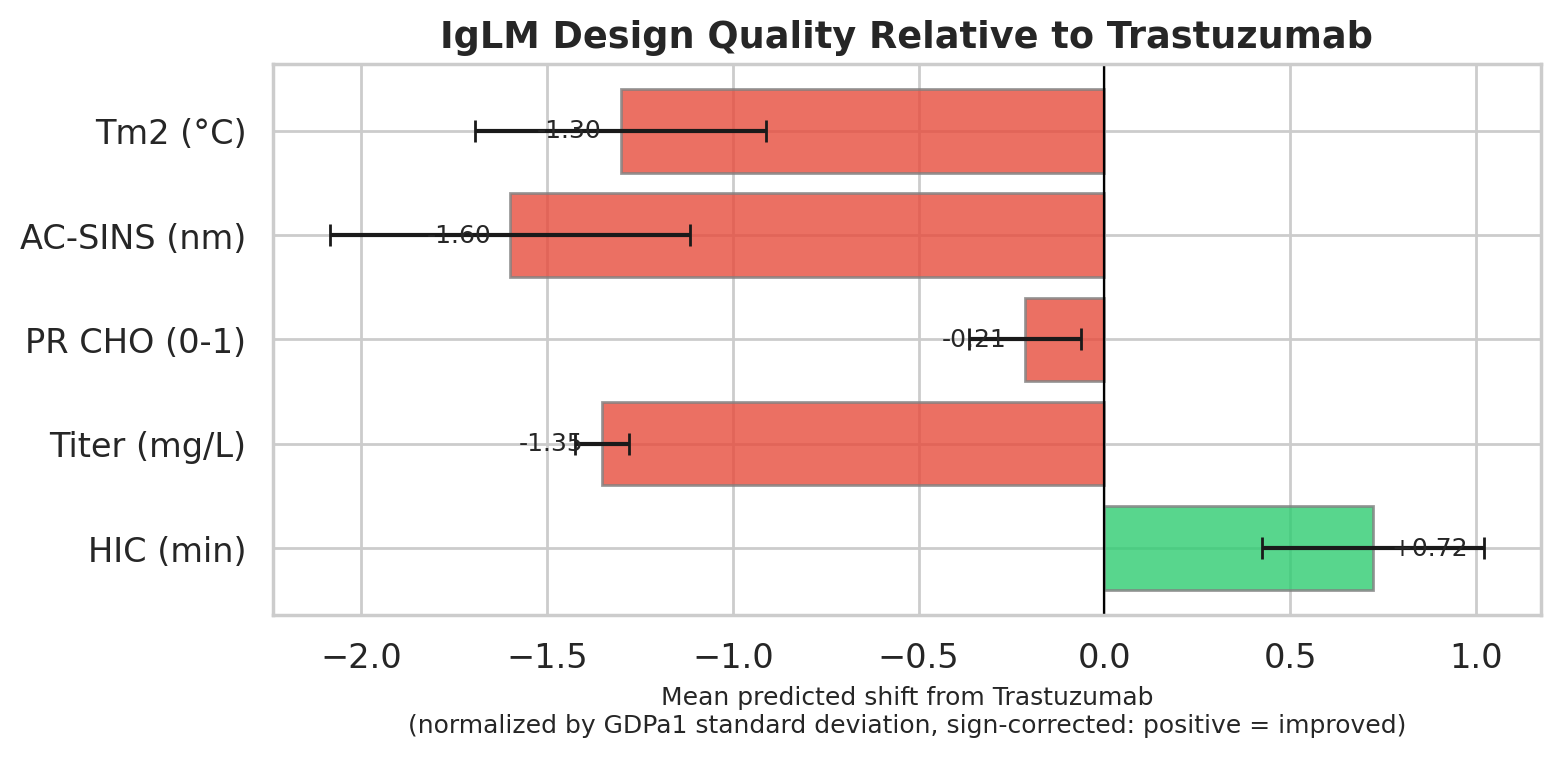}
\caption{Developability delta of 100 IgLM-generated designs relative to the trastuzumab reference, normalized by GDPa1 training set standard deviation (sign-corrected: positive = improvement). Error bars indicate inter-design standard deviation. Only HIC shows consistent improvement, PR\_CHO clusters near zero (matching trastuzumab's floor); and all other properties degrade, with the largest deficits on AC-SINS and Titer --- the two properties requiring inter-chain reasoning.}
\label{fig:iglm_delta}
\end{figure}

\section{Reproducibility}
\label{app:reproducibility}

\begin{itemize}
    \item \textbf{Dataset}: GDPa1 benchmark~\citep{Arsiwala2025}, 242 therapeutic IgGs, 5 assays.
    \item \textbf{Splits}: 5-fold cluster-stratified cross-validation with IgG isotype stratification. Clustering ensures sequence-similar antibodies are separated across folds.
    \item \textbf{Metric}: Spearman rank correlation ($\rho$), averaged across folds.
    \item \textbf{Baselines}: All baseline results reported from~\citet{Arsiwala2025} under identical CV splits.
    \item \textbf{Compute}: All experiments run on a single NVIDIA GPU. Encoder embeddings are precomputed and frozen. Decoder training takes approximately 5 minutes per fold per property. Full search campaign exploring architectural choices and a range of hyperparameters: $\sim$200 configurations per property $\times$ 5-fold CV.
    \item \textbf{Code}: CrossAbSense is available open-source at \url{https://github.com/SimonCrouzet/CrossAbSense}, including all training scripts, sweep configurations, and evaluation code.
\end{itemize}

\end{document}